%% file: bad1808_la.tex
\documentclass[twocolumn,showpacs,aps,prl,superscriptaddress]{revtex4}

\usepackage{graphicx}
\usepackage{dcolumn}
\usepackage{amsmath}
\usepackage{xspace}

\input pubboard/babarsym.tex

\input definitions.tex

\newcommand{\BABARPubYear}    {08}
\newcommand{\BABARPubNumber}  {008}

\newcommand{\SLACPubNumber} {13198}
\newcommand{\LANLNumber} {0804.1208}

\begin{document}

\begin{flushleft}
\babar-PUB-\BABARPubYear/\BABARPubNumber\\
SLAC-PUB-\SLACPubNumber\\
arXiv: \LANLNumber~$[{\rm hep-ex}]$\\
\end{flushleft}

\title{\Large \boldmath Study of \B-meson decays to $\etac K^{(*)}$, $\etactwos K^{(*)}$ and $\etac\gamma K^{(*)}$}

\date{\today}
\input pubboard/authors_feb2008.tex

\begin{abstract}
We study two-body \B-meson decays to a charmonium state (\etac, \etactwos\ or \hc) and a $\Kp$ or $\Kstarz(892)$ meson using a sample of 349~\invfb\ of data collected with the \babar\ detector at the \pep2\ asymmetric-energy \B\ Factory at SLAC. We measure $\BR(B^0\to\etac\Kstarz)=(5.7\pm0.6 (\mathrm{stat})\pm0.9 (\mathrm{syst}))\times 10^{-4}$, $\BR(B^0\to\etactwos \Kstarz)<3.9\times 10^{-4}$, $\BR(B^+\to\hc \Kp)\times \BR(\hc\to\etac\gamma)<4.8\times 10^{-5}$ and $\BR(B^0\to\hc \Kstarz)\times \BR(\hc\to\etac\gamma)<2.2\times 10^{-4}$ at the $90\%$ C.L., and $\BR(\etactwos\to\kkpi)=(1.9\pm0.4 (\mathrm{stat}) \pm1.1 (\mathrm{syst}))\%$. We also measure the mass and width of the \etac\ meson to be $m(\etac)=(2985.8\pm 1.5(\mathrm{stat}) \pm 3.1 (\mathrm{syst}))~\mevcc$ and $\Gamma(\etac)=(36.3^{+3.7}_{-3.6} (\mathrm{stat}) \pm 4.4 (\mathrm{syst}))~\mev$.
\vfill
\end{abstract}


\maketitle

In the simplest approximation, $B$ decays to a charmonium state and a $K$ or $K^*$ meson  arise from the quark-level process \mbox{$b \to \ccbar s$} and have been observed to occur with large rates~\cite{ref:pdg2006}. However several decay modes are still poorly known, particularly in the case of singlet states such as \etac\ and \hc.  A better knowledge of the relative abundances of the decay to the various charmonium states allows a deeper understanding of the underlying strong processes and tests of the predictions of models such as non-relativistic QCD~\cite{ref:NRQCD}. In non-relativistic QCD, the $B$ decay rates to all P-wave states of charmonium, $\chi_{cJ}$ ($J=0,1,2$) and \hc, do not vanish and are foreseen to be comparable in magnitude. 

In this document, we study $B$-meson decays to $(\kkpi)\Kp$, $(\kkpi)\Kstarz$, $\etac\gamma \Kp$ and $\etac\gamma \Kstarz$, from which we measure the branching fractions for the following decay modes: $B^0\to\etac\Kstarz$, $B^0\to\etactwos \Kstarz$, $B^0\to \hc \Kstarz$, $B^+\to \hc \Kp$~\cite{ref:cc_note}, and $\etactwos\to\kkpi$. We also measure the mass and width of the \etac\ meson. The \hc\ meson has recently been discovered by the CLEO Collaboration as a narrow peak at $3524.4\pm0.7~\mevcc$ in the $\etac\gamma$ invariant mass distribution in $\psi(2S)\to\etac\gamma\piz$ decays~\cite{ref:CLEO_1P1}, and this observation was confirmed by the E835 Collaboration~\cite{ref:E835_1P1}. The \etactwos\ state was discovered by the Belle Collaboration in $B$ decays to $(\kskpi) K$~\cite{ref:Belle_etapc1}, and subsequently observed in the processes $\gaga\to\etactwos\to\kskpi$ and $\epem\to\jpsi\etactwos$; its mass is $(3637\pm4)~\mevcc$ and its width is $(14\pm7)~\mev$~\cite{ref:pdg2006}. No branching fraction for any \etactwos\ decay mode is yet listed by the Particle Data Group~\cite{ref:pdg2006}. Using the measured value for $\Gamma(\etactwos\to\gaga)\times\BR(\etactwos \to \kkpi)$~\cite{ref:CLEO_etac2S}, a measurement of $\BR(\etactwos \to \kkpi)$ can be used as an input to derive $\Gamma(\etactwos\to\gaga)$, a quantity calculable in a theoretically clean way within the conventional framework of QCD: calculations that assume $\BR(\etactwos \to \kkpi)=\BR(\etac \to \kkpi)$ lead to values of $\Gamma(\etactwos\to\gaga)$ smaller than expectations, pointing to a possible anomaly in the \etactwos\ decay~\cite{ref:pham}. The branching fraction of $B^0\to\etac\Kstarz$ is currently known with a $40\%$ uncertainty, $(1.2\pm0.5)\times 10^{-3}$~\cite{ref:etackstar_note}, while $B$ decays to $\etactwos K^*$ and $\hc K^{(*)}$ have never been observed. The Belle Collaboration studied the decay $B^+\to\hc \Kp$ with $\hc\to\etac\gamma$ and reported $\BR(B^+\to \etac\gamma \Kp) < 3.8\times 10^{-5}$ at the $90\%$ C.L. for an invariant mass of the $\etac\gamma$ pair in the range [3.47,3.57]~\gevcc~\cite{ref:belle_exclhc}. This limit is comparable to analogous limits for \chictwo\ but significantly smaller than the measured branching fractions for $B$ decays to \etac, \jpsi, \chiczero\ or \chicone\ and a kaon~\cite{ref:pdg2006}. No other $B^+$ or $B^0$ decay modes with \hc\ have yet been studied. The mass and width of the \etac\ are important parameters in models of the charmonium spectrum~\cite{ref:potmodels}: the hyperfine separation $(\etac,\jpsi)$ is directly related to the spin-spin interaction. The \etac\ mass and width measurements reported in the literature~\cite{ref:pdg2006} are often in poor agreement with one another. The listed world average for the mass is $(2979.8\pm1.2)~\mevcc$, with measurements ranging from 2969 to 2984~\mevcc, and for the width it is $(26.5\pm3.5)~\mev$ with values ranging from 7 to 48~\mev. 

In this analysis we reconstruct the \etac\ and \etactwos\ in the \kskpi\ and $\Kp\Km\piz$ decay modes, the \hc\ in its decay to $\etac\gamma$, the \KS\ in the mode $\KS\to\pipi$ and the \Kstarz\ in $\Kstarz\to \Kp\pi^-$. The \kskpi\ and $\Kp\Km\piz$ final states are chosen because they are among the easier \etac\ decay modes to reconstruct and have a rather large branching fraction, $\BR(\etac\to K\bar{K}\pi)=(7.0\pm1.2)\%$~\cite{ref:pdg2006}. For the \etactwos, the \kskpi\ mode is the only decay observed so far. The $\etac\gamma$ decay of the \hc\ is chosen because it is expected to comprise about half of the total \hc\ decay width~\cite{ref:NRQCD}. For decays with \etac\ and \hc, we measure ratios of branching fractions with respect to $\BR(B^+\to\etac\Kp)=(9.1\pm1.3)\times 10^{-4}$~\cite{ref:pdg2006}, to cancel the $17\%$ uncertainty on $\BR(\etac\to\kkpi)$. Similarly, we measure the ratio $\BR(B^0\to\etactwos \Kstarz)/\BR(B^+\to\etactwos \Kp)$, to cancel the unknown branching fraction of $\etactwos\to\kkpi$. 

The data used in this analysis were collected with the \babar\ detector at the \pep2\ \epem\ storage rings, and correspond to 349~\invfb\ of integrated luminosity collected at  the \FourS\ resonance, comprising $384$ million \BB\ pairs. The \babar\ detector is described elsewhere~\cite{ref:babar}. We make use of Monte Carlo (MC) simulations based on GEANT4~\cite{ref:geant4}.

The event selection is optimized by maximizing the quantity $N_S/\sqrt{N_S+N_B}$, where $N_S$ ($N_B$) represents the number of signal (background) candidates surviving the selection. $N_S$ is estimated from samples of simulated events of $B\to\etac K^{(*)}, \etac\to\kkpi$ decays for $B\to(\kkpi)K^{(*)}$, and $B\to\hc K^{(*)}, \hc\to\etac\gamma, \etac\to\kkpi$ for $B\to\etac\gamma K^{(*)}$. $N_B$ is estimated from signal sidebands on data, defined by the signal candidates with reconstructed \epem\ center-of-mass energy farther than 3 standard deviations from the expectation in \epem\ collisions at the \FourS\ peak. Simulated signal events and data are normalized to each other using the available measurements for $B$ decays to \etac\ and assuming $\BR(B\to\hc K^{(*)})=1\times 10^{-5}$.

We select events with \BB\ pairs by requiring at least four charged tracks, the ratio of the second to the zeroth order Fox-Wolfram moment~\cite{ref:FoxWolfram} to be less than 0.2, and the total energy of all the charged and neutral particles to be greater than 4.5~\gev. 

Charged pion and kaon candidates are reconstructed tracks having at least 12 hits in the drift chamber, a transverse momentum with respect to the beam direction larger than 100 \mevc, and a distance of closest approach to the beam spot smaller than 1.5~cm in the plane transverse to the beam axis and 10~cm along the beam axis. We use particle identification provided by measurements of the energy loss in the tracking devices and the Cherenkov detector.
A $\Kstarz$ candidate is formed from a pair of oppositely charged kaon and pion candidates originating from a common vertex and having an invariant mass within 60~\mevcc\ of the nominal $\Kstarz$ mass~\cite{ref:pdg2006}.

Photon candidates are energy deposits in the electromagnetic calorimeter that are not associated with charged tracks, having energy greater than 100~\mev\ and a shower shape consistent with that of a photon. A \mbox{\piz\to\gaga} candidate is formed from a pair of photon candidates with invariant mass in the range [115,150]~\mevcc\ and energy greater than 400~\mev. These candidates are constrained to the nominal \piz\ mass~\cite{ref:pdg2006}.

A \KS\to\pipi\ candidate is formed from a pair of oppositely charged tracks originating from a common vertex and having an invariant mass within 20~\mevcc\ of the $K^0$ mass. Its  measured decay-length significance is required to exceed three standard deviations. The candidate is constrained to the nominal $K^0$ mass~\cite{ref:pdg2006}.

The $B^{+,0}\to(\kkpi) K^{+,*0}$ candidates are formed by pairing a \Kstarz\ or $\Kp$ candidate, referred to as the primary kaon, and a \kskpi\ or $\Kp\Km\piz$ combination with invariant mass above $2.75~\gevcc$ to include the whole charmonium region.  The $B^{+,0}\to\etac\gamma K^{+,*0}$ candidates are formed by combining a \Kstarz\ or $\Kp$ candidate, a photon with energy exceeding 250~\mev, and  a \kskpi\ or $\Kp\Km\piz$ combination with invariant mass consistent with the \etac\ mass. We perform a vertex fit to the \B\ candidates and require the $\chi^2$ probability to exceed 0.002. We define two kinematic variables: the beam-energy substituted mass, $\mes=\sqrt{E^{2}_{\mathrm{beam}}-p^{2}_B}$ and $\Delta E=E_B-E_{\mathrm{beam}}$, where $p_B$ ($E_B$) is the reconstructed \B\ momentum (energy) and $E_{\mathrm{beam}}$ is the beam energy, in the $e^+e^-$ center-of-mass (c.m.) frame. $B$ candidates are retained if they have \mes\ greater than 5.2~\gevcc\ and $\Delta E$ within [$-$24,30], [$-$40,30], [$-$34,30], and [$-$40,30]~\mev\ for the $\kskpi K^{*0,+}$, $\Kp\Km\piz K^{*0,+}$, $\kskpi\gamma K^{*0,+}$, and $\Kp\Km\piz\gamma K^{*0,+}$ combinations, respectively. $B$ mesons produced in the process $\FourS\to\BB$ follow a $\sin^2\theta_B$ distribution, where $\theta_B$ is the polar angle of the \B\ candidate momentum vector in the \epem\ c.m. frame: we require $|\cos\theta_B|<0.9$.

To suppress background, $\Kp\pi^-$, $\Kp\Km$, $\Kp\KS$, $\KS\pi^-$ and $\Kp\pi^-\pi^+$ combinations with invariant masses within 30~\mevcc\ of the $D^0$, $D_s$ and $D^+$ meson masses~\cite{ref:pdg2006} are excluded when forming $B$ candidates. We also remove $\Kp\Km$ pairs containing a primary kaon where the invariant mass of the pair is within 30~\mevcc\ of the $\phi$ meson mass~\cite{ref:pdg2006}.

In events where more than one $B$ candidate survives the selection, the one with the smallest $|\Delta E|$ is retained. In cases of multiple $B$ candidates composed from the same final state particles, and thus having the same value of $|\Delta E|$, we retain the one for which the primary kaon  has the largest momentum in the \epem\ c.m. frame.

The samples surviving the selection include a signal component, a combinatorial background component given by random combinations of tracks and neutral clusters both from \BB\ and continuum events $\epem\to\qqbar$ ($q=u,d,s,c$), and a component due to $B$ decays with a similar final state to the signal. As opposed to the combinatorial background, such ``peaking backgrounds'' exhibit the same distribution as signal events in \mes\ and $\Delta E$, but their $\kkpi(\gamma)$ invariant-mass distribution ($m_X$) is different. The signal content in data is therefore obtained by means of a maximum likelihood fit to $m_X$ for all candidates having \mes\ in the signal region $[5.274,5.284]$~\gevcc, after subtracting the combinatorial background. The $m_X$ distribution for the combinatorial background events is obtained by extrapolating into the \mes\ signal region the $m_X$ distribution measured in the \mes\ sideband, defined by $5.20 < \mes < 5.26~\gevcc$. The correlation between $m_X$ and \mes\ is found to be negligible in the relevant regions.  A binned fit is then performed on the \mes-sideband-subtracted \mX\ distribution. 

\begin{figure}[!tb]
  \begin{center} 
    \includegraphics[scale=0.2]{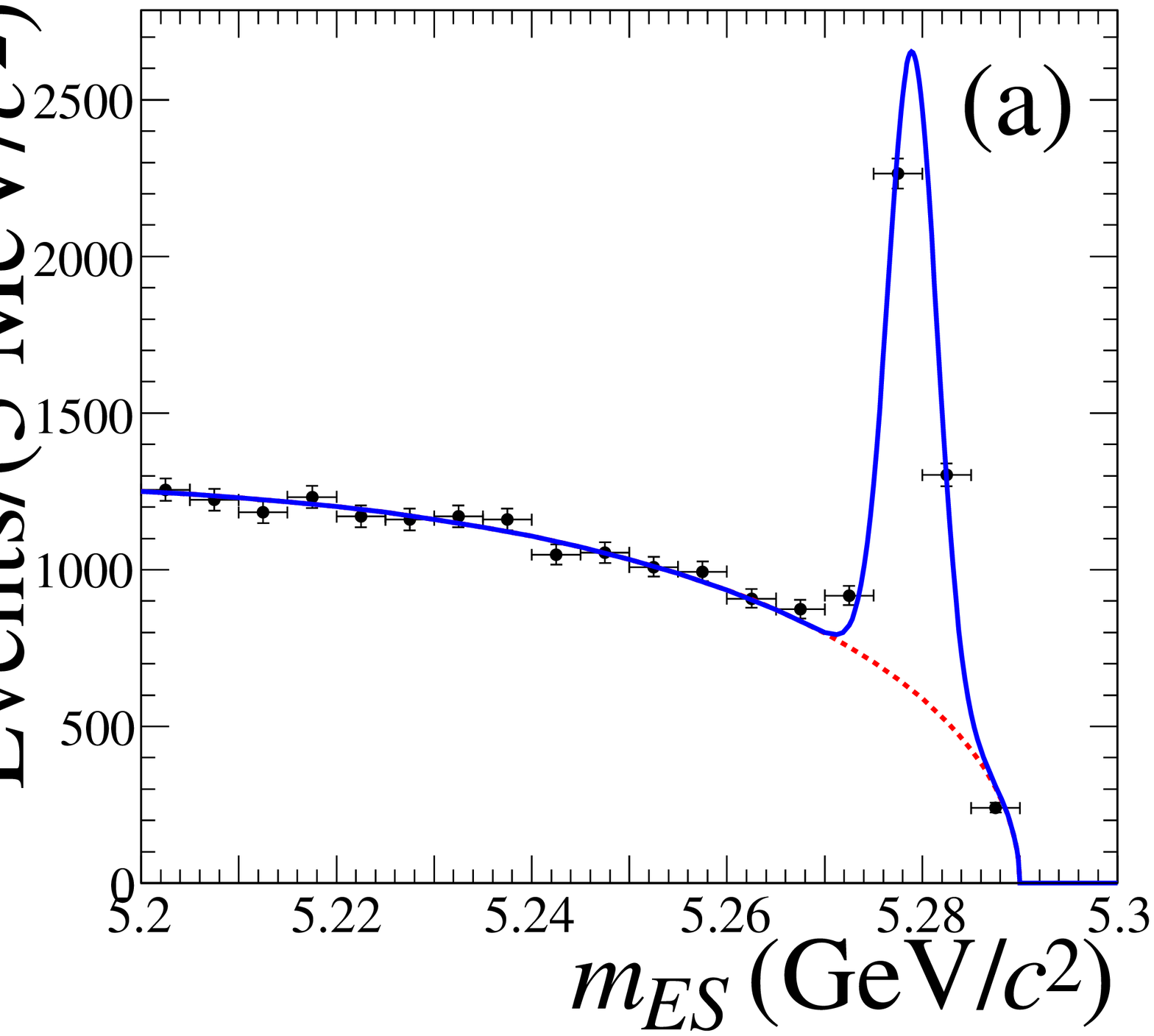}
    \includegraphics[scale=0.2]{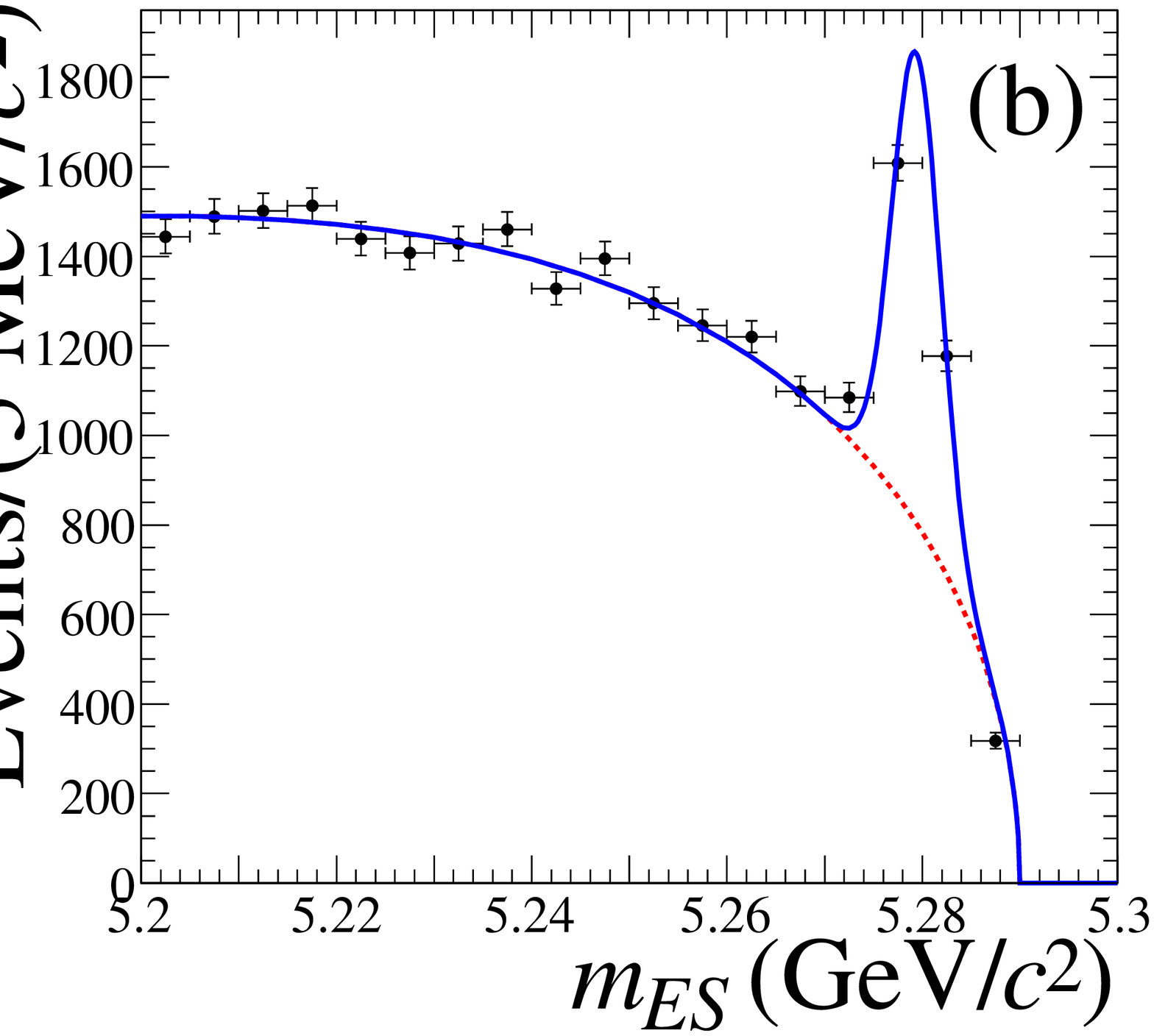}
    \includegraphics[scale=0.2]{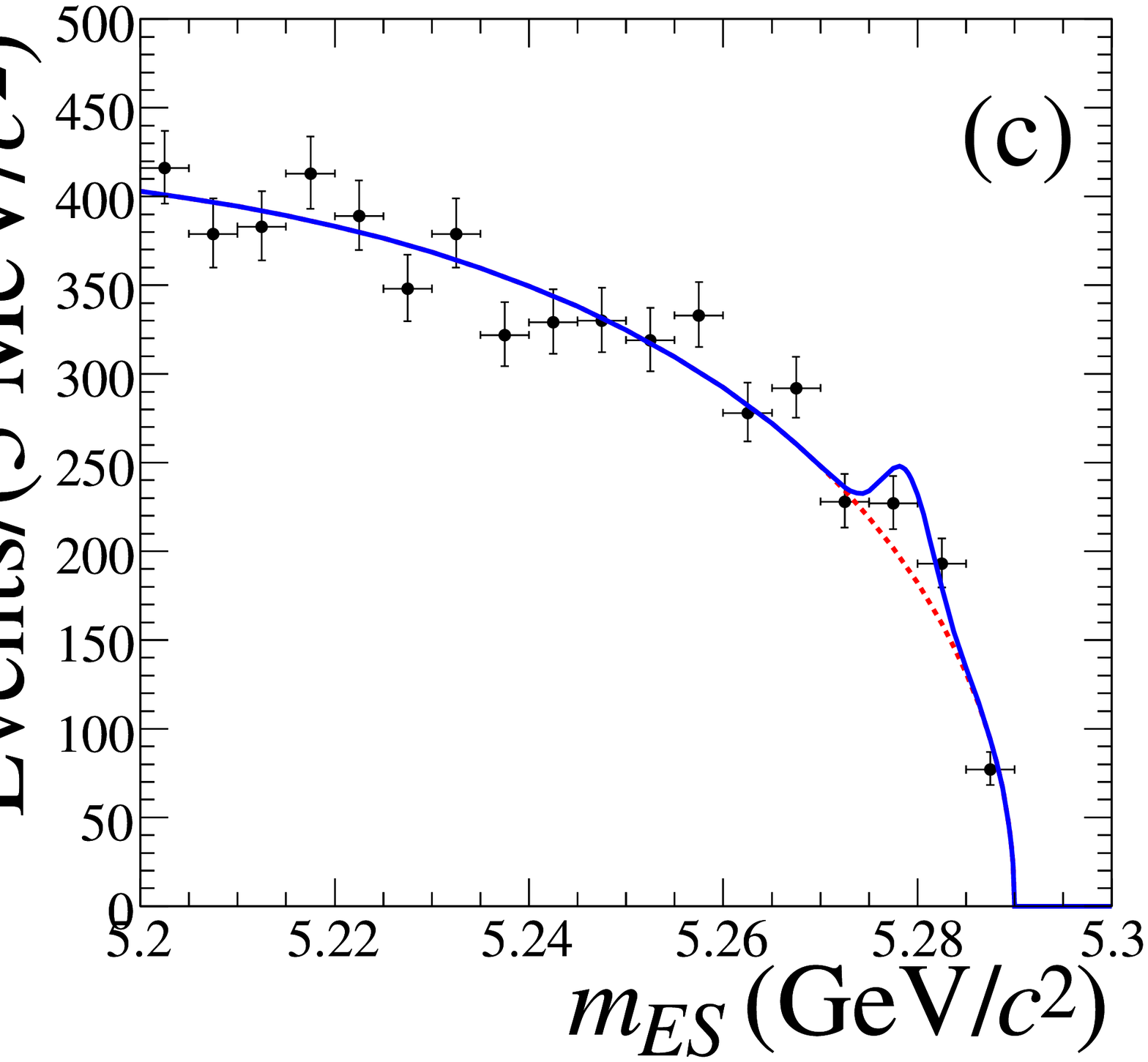}
    \includegraphics[scale=0.2]{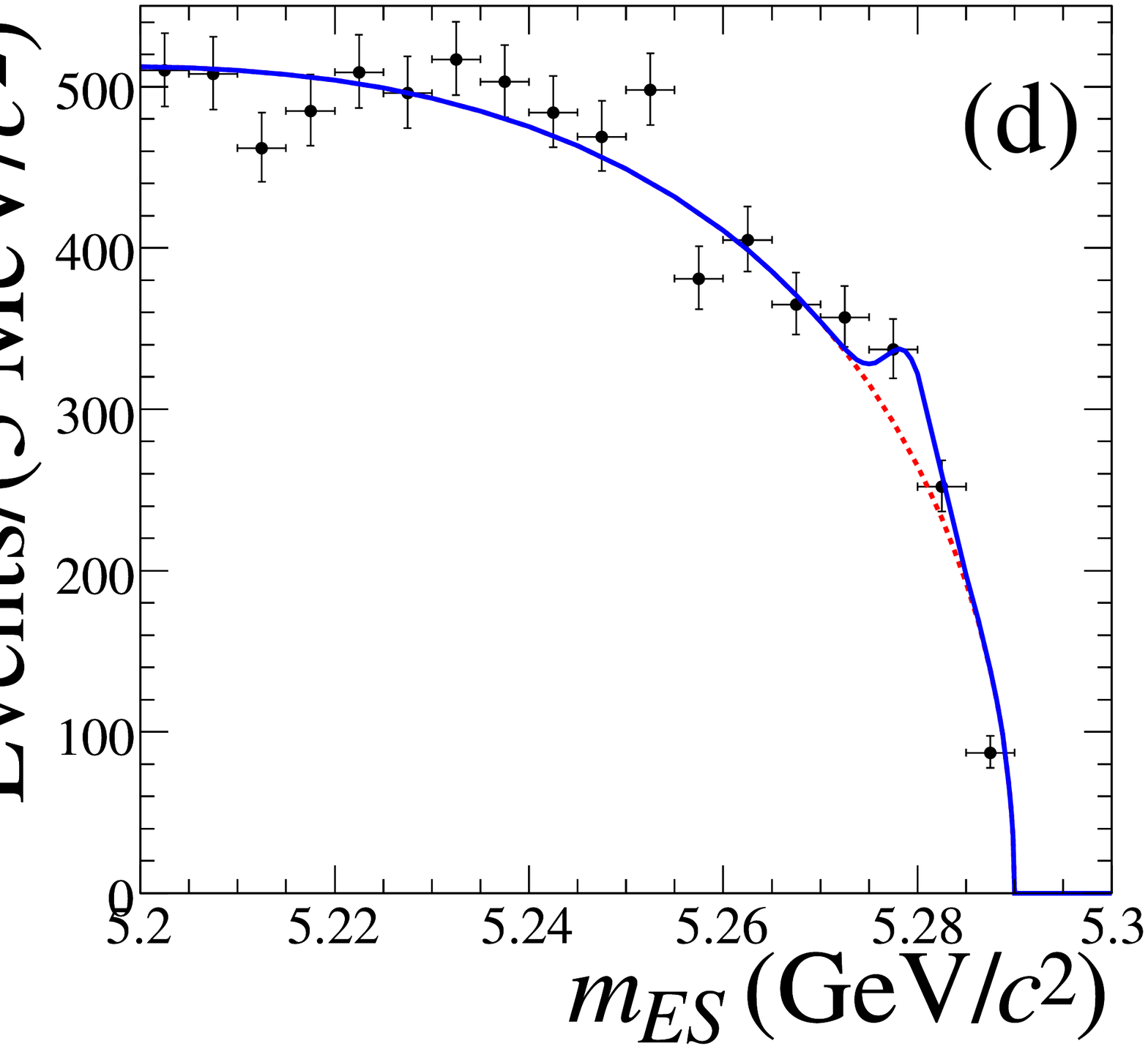}
\caption{\footnotesize{The \mes\ distributions for (a) $B^+\to (K\bar{K}\pi) K^{+}$, (b) $B^0\to (K\bar{K}\pi) \Kstarz$, (c) $\Bp\to\etac\gamma\Kp$ and (d) $B^0\to\etac\gamma \Kstarz$ candidates; points with error bars are data, the solid line represents the result of the fit described in the text, and the dotted line represents the ARGUS background parameterization. No appreciable $B$ component, either signal or peaking background, is observed for the $\Bp\to\etac\gamma\Kp$ and $B^0\to\etac\gamma \Kstarz$ cases.}\label{fig:signalExtr_mes}}
  \end{center}
\end{figure}

To estimate the background we perform an unbinned maximum likelihood fit to the \mes\ distribution as follows. The \B\ component, accounting for the sum of signal and peaking background, is modelled by a Gaussian function whose width is taken from the simulation and whose mean is fixed to the $B$-meson mass~\cite{ref:pdg2006}. The \mes\ distribution of the combinatorial background is represented by an ARGUS function~\cite{ref:Argus}. The total number of events and the exponent of the ARGUS function are left free in the fit. The spectrum for candidates in the \mes\ sideband is normalized to the \mes\ signal window by using the integrals of the ARGUS component in the two regions (Fig.~\ref{fig:signalExtr_mes}).

\begin{figure}[!htb]
  \begin{center} 
    \includegraphics[scale=0.35]{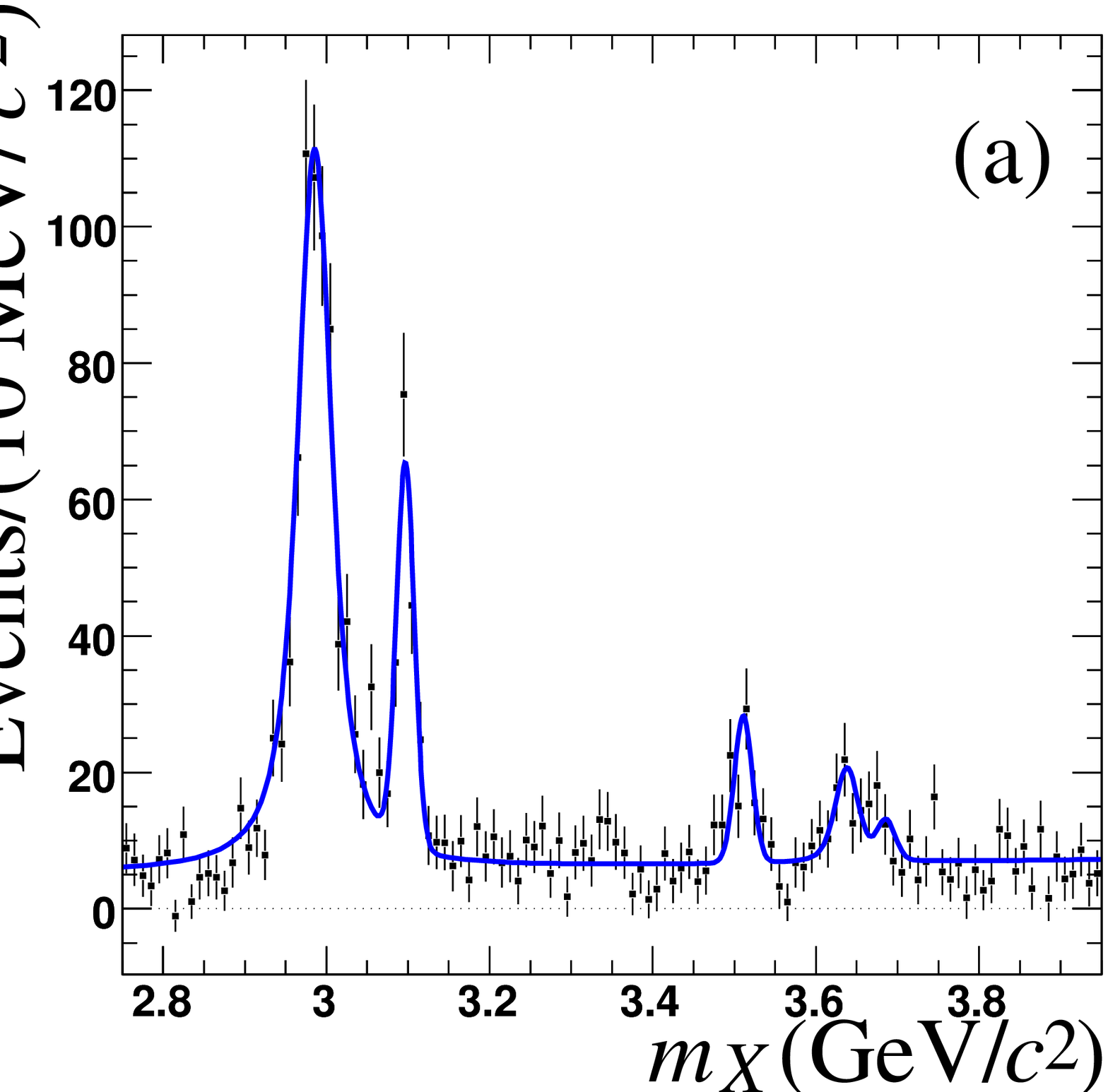} 
    \includegraphics[scale=0.35]{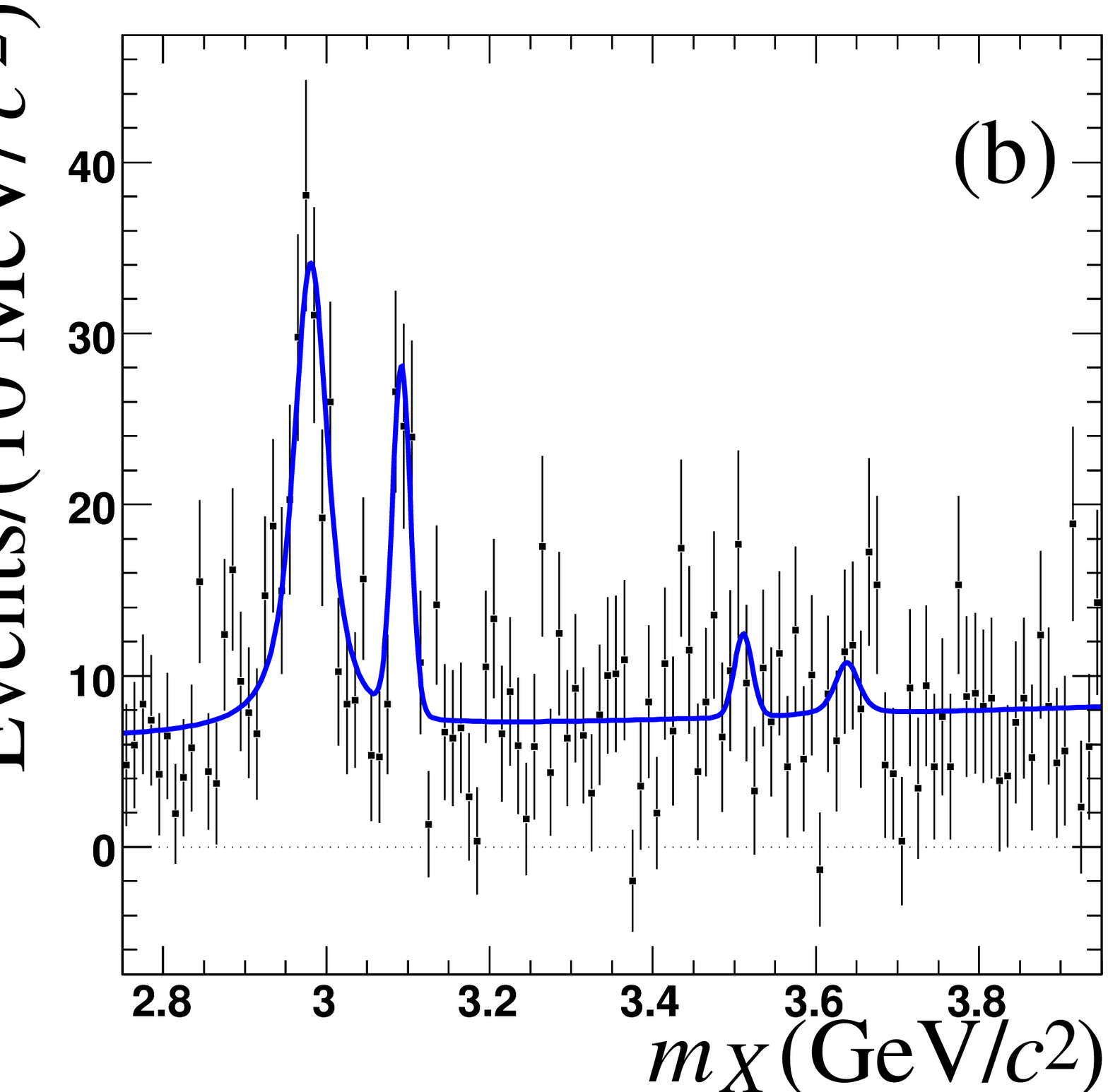} 
\caption{\footnotesize{Fit result (solid line) superimposed on the \mes-sideband-subtracted \mX\ distribution (points with error bars) for (a) $B^+\to (K\bar{K}\pi) K^{+}$ and (b) $B^0\to (K\bar{K}\pi) \Kstarz$.}\label{fig:signalExtr_mx_etac}}
  \end{center}
\end{figure}

The \mX\ distribution for $\Bp\to(\kkpi)\Kp$ and $B^0\to(\kkpi)\Kstarz$ is shown in Fig.~\ref{fig:signalExtr_mx_etac}, after subtraction of the \mes\ sideband background. The two samples are simultaneously fitted to the sum of an \etac, an \etactwos, a \jpsi, a \chicone\ and a \psitwos, and a background component accounted for by first-order polynomials. The \etac\ and \etactwos\ peaks are modelled by a non-relativistic Breit-Wigner convolved with a Gaussian function, the others by Gaussians. The masses of \chicone, \etactwos\ and \psitwos\ and the width of the \etactwos\ are fixed to the world average values~\cite{ref:pdg2006}. To reduce systematic uncertainties on the \etac\ mass measurement from potential distortion effects in data shifting the peak positions, in the fit  we float the mass of the \jpsi\ and fit for the mass difference between \jpsi\ and \etac. We also float the width of the \etac; the mass resolutions, modelled by the widths of the Gaussian functions, separately for the \kskpi\ and $\Kp\Km\piz$ modes; the coefficients of the background polynomial functions and the number of signal and background events. The fit extends over the \mX\ range [2.75,3.95]~\gevcc. No component is included for other charmonium states such as \chiczero, \hc\ and \chictwo, since they have not been observed to decay to \kkpi\ and/or in $B$ decays. Table~\ref{tab:yields_etac} summarizes the numbers of events found by the fit, separately for the $\Bp\to(\kkpi)\Kp$ and $B^0\to(\kkpi)\Kstarz$ samples. The $\chi^2$ of the fit divided by the number of degrees of freedom ($N_{Dof}$) is 1.2. The mass resolution is determined by the fit to be $9\pm1$~\mevcc\ and $20\pm9$~\mevcc\ for \kskpi\ and $\Kp\Km\piz$, respectively. The mass of the \jpsi\ is found to be $3096.4\pm1.0~\mevcc$, the mass difference between \jpsi\ and \etac\ $111.1\pm1.5~\mevcc$, and the \etac\ width $36.3^{+3.7}_{-3.6}~\mev$. Using $m(\jpsi)=3096.916\pm0.011~\mevcc$ from~\cite{ref:pdg2006}, we derive $m(\etac)=2985.8\pm1.5~\mevcc$.
\begin{table}[!htb]
\caption{Numbers of \etac, \jpsi, \chicone, \etactwos\ and \psitwos\ events obtained from the fit described in the text with statistical errors only.}
\begin{ruledtabular}
\begin{tabular}{cccc} 
\rule[-1mm]{0mm}{4ex} & $B^+\to (\kkpi) \Kp$     & $B^0\to (\kkpi) \Kstarz$ \\
\hline
$N_{\etac}$               & $732\pm 27$ & $189\pm 18$\\
$N_{\jpsi}$               & $154\pm 15$ & $56\pm 9$\\
$N_{\chicone}$            & $59\pm 10$  & $13\pm 7$\\
$N_{\etactwos}$           & $59\pm 12$  & $13\pm 9$\\
$N_{\psitwos}$            & $15\pm 8$   & $0\pm 4$\\
\end{tabular}
\end{ruledtabular}
\label{tab:yields_etac}
\end{table}

\begin{figure}[!htb]
  \begin{center} 
    \includegraphics[scale=0.35]{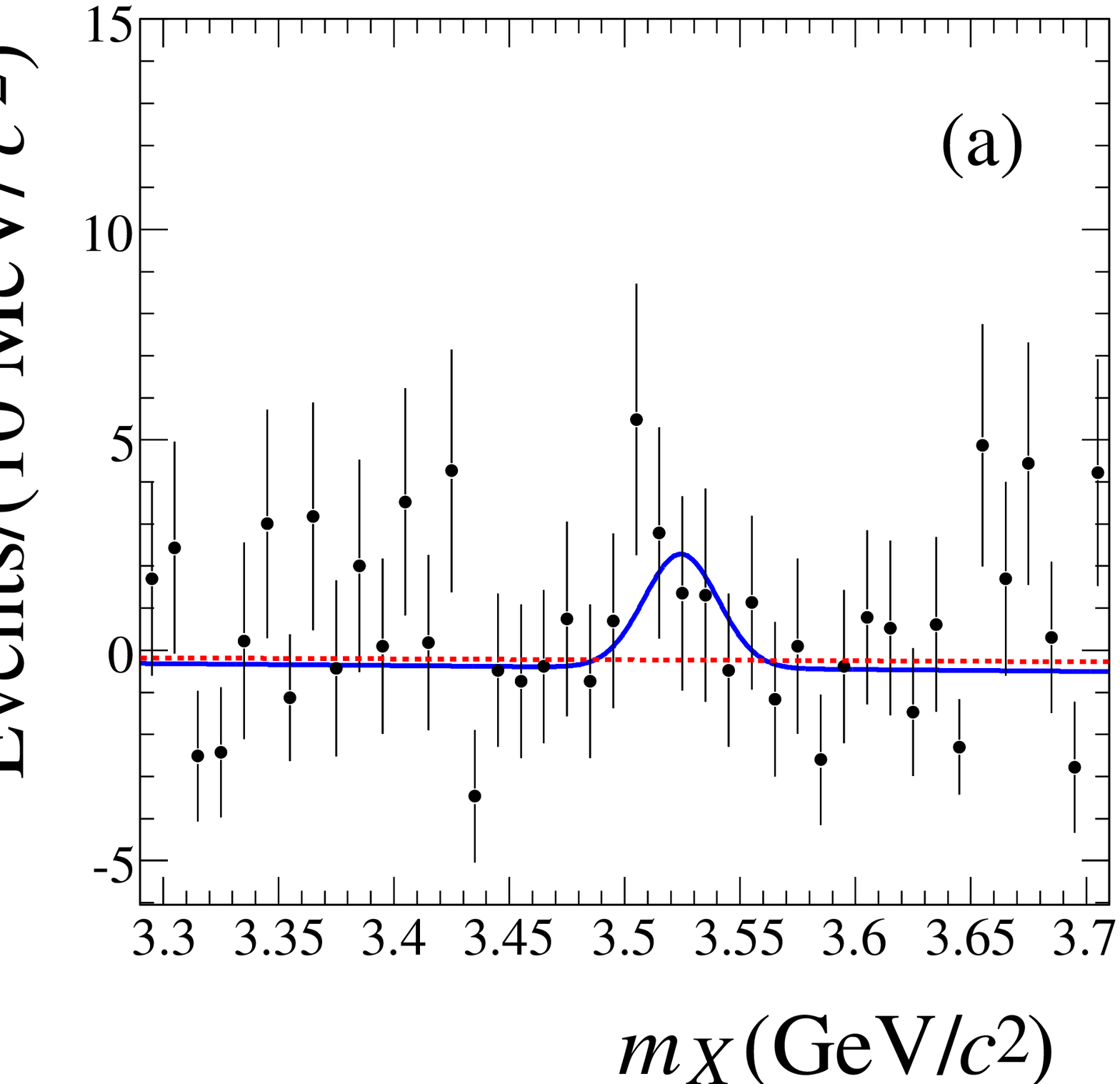}
    \includegraphics[scale=0.35]{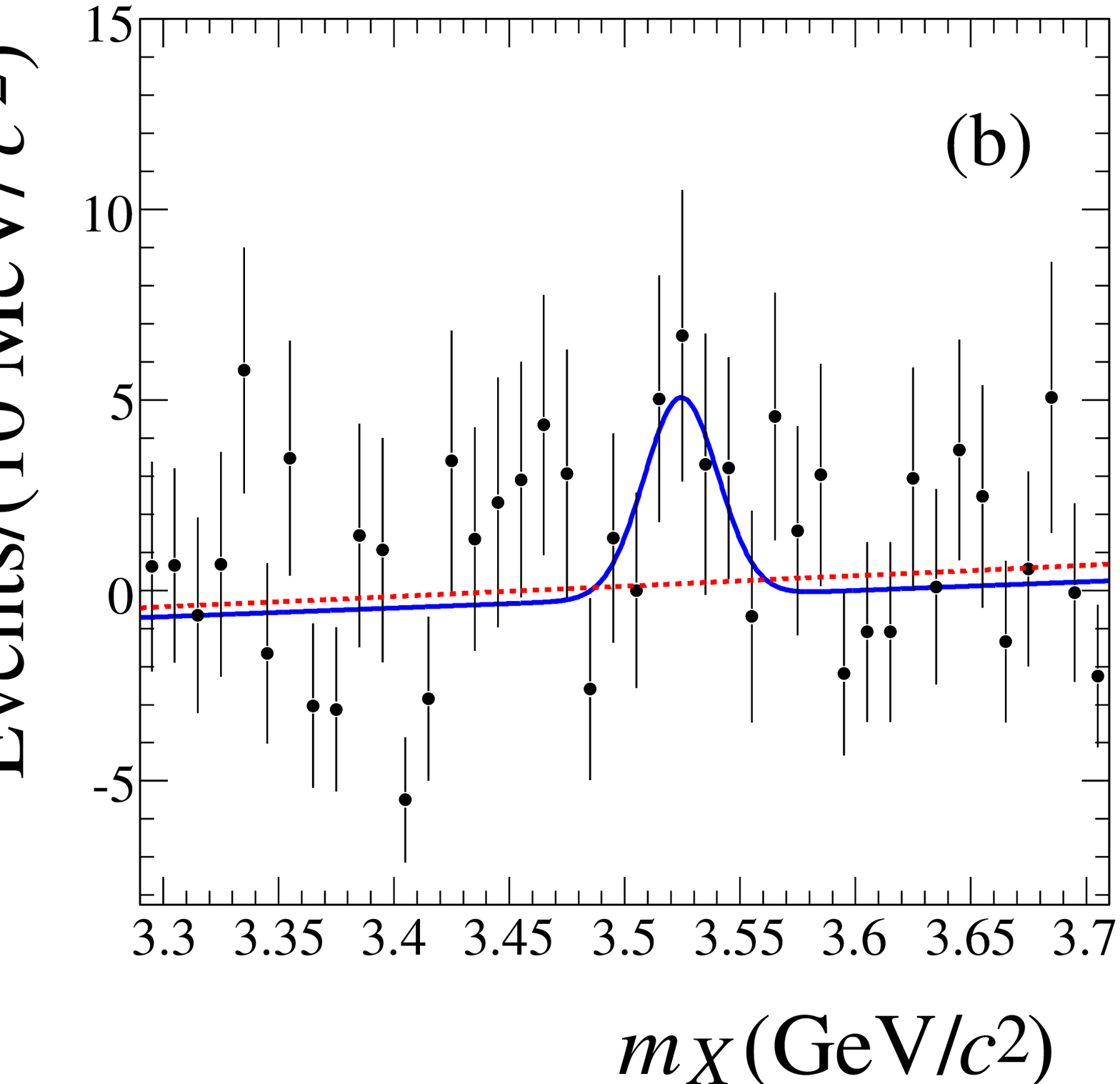}
\caption{\footnotesize{Fit result (solid line) superimposed on the \mes-sideband-subtracted \mX\ distribution (points with error bars) for (a) $\Bp\to\etac\gamma\Kp$ and (b) $B^0\to\etac\gamma \Kstarz$. No significant \hc\ signal is evident. The dashed line is the result of the fit with no signal component.
}\label{fig:signalExtr_mx_hc}}
  \end{center}
\end{figure}

In the case of $\Bp\to\etac\gamma\Kp$ and $B^0\to\etac\gamma \Kstarz$ (Fig.~\ref{fig:signalExtr_mx_hc}), the \mes-sideband-subtracted $m_X$ distribution is fitted to the sum of  an \hc\ signal modelled by a Gaussian, and a background represented by a first-order polynomial. The mass of the \hc\ is fixed to the CLEO measurement, 3524~\mevcc~\cite{ref:CLEO_1P1}. The Gaussian resolution is fixed to the value determined from MC events, $16~\mevcc$~\cite{ref:hc_simul}. In the fit, the numbers of signal and background events are left free. The fit is performed over the \mX\ range [3.3,3.7]~\gevcc. It yields $11\pm6$ and $21\pm8$ \hc\ candidates with a $\chi^2/N_{DoF}$ of $41/39$ and $42/39$ for the $B^+$ and $B^0$ yields, respectively. 
 
The stability of the fit results is verified for various configurations of the fitting conditions. For  $B\to (K\bar{K}\pi) K^{(*)}$, we perform the fits with and without components for \chiczero, \chictwo, \hc\ and \psitwos\ in various combinations. The values for the signal yields and the floated parameters returned by these fits are consistent with the nominal configuration. We validate the fit procedure using a MC technique: we simulate a number of experiments by randomly generating samples of events distributed in \mX\ according to the models used in the fit. The number of events generated is equal to the number of events in the corresponding real data sample. The parameters of the distributions are set to their fixed or fitted values. The fit is repeated under the same conditions as used on real data. The numbers of signal and background events are distributed as expected. The robustness of the fit is tested on simulated events by varying the numbers of signal and background events input, including the null result. The numbers of events returned by the fit are consistent with the inputs for all cases. As additional cross-checks, we verify that the observed numbers of \jpsi, \chicone\ and \psitwos\ candidates in the data agree with the expectations.

We evaluate systematic uncertainties on the numbers of signal candidates and the mass and width determination by individually varying the parameters that are fixed in the fits by $\pm 1$ standard deviation from their nominal values. We also estimate the systematic uncertainties that arise from a different choice of binning, fit range, and background parameterization. For $\Bp\to(\kkpi)\Kp$ and $B^0\to(\kkpi)\Kstarz$, where the mass resolutions are floated, we estimate an additional systematic uncertainty by taking the variations with respect to a fit performed by fixing the mass resolutions to the values determined from the simulation, 8~\mevcc\ and 19~\mevcc\ for \kskpi\ and $\Kp\Km\piz$, respectively. The large natural widths of the \etac\ and \etactwos\ introduce the possibility of interference effects with non-resonant $B$ decays to the same final state particles. This can modify the $m_X$ distribution with respect to the one used in the fit. The fit is repeated including an interference term between the \etac\ and the background in the fitting functions. The amplitude and phase of the interference term are left free in the fit. The variation of the \etac\ yield with respect to the nominal fit is taken as an estimate of the systematic error due to neglecting interference effects. A similar approach is undertaken for \etactwos.
Summing in quadrature all the contributions, the total systematic uncertainty on the signal yield determination is $6\%$, $3\%$, $25\%$, $18\%$, $25\%$ and $23\%$  for $B^{+}\to\etac\Kp$, $B^{0}\to\etac\Kstarz$, $B^{+}\to\hc\Kp$, $B^{0}\to\hc\Kstarz$, $B^{+}\to\etactwos\Kp$ and $B^{0}\to\etactwos\Kstarz$, respectively, and the total systematic uncertainties on the \etac\ mass and width are 3.1~\mevcc\ and 4.4~\mev, respectively.

The selection efficiency for $B^+\to\etac\Kp$ is $6 \%$. The ratios of the selection efficiencies with respect to $B^+\to\etac\Kp$, estimated by using simulated events, are, including systematic uncertainties, $0.64\pm 0.01$, $0.51\pm 0.01$, $0.29\pm0.02$, $0.84\pm0.01$ and $0.54\pm0.01$ for $B^0\to\etac\Kstarz$, $B^+\to\hc K^{+}$, $B^0\to\hc \Kstarz$, $B^{+}\to\etactwos K^{+}$ and $B^{0}\to\etactwos \Kstarz$, respectively. Most uncertainties on the efficiencies cancel out in the ratios because of the similar final states. The remaining uncertainties are mainly due to differences between real data and simulation in the photon reconstruction as estimated from photon control samples from data ($1.8\%$), and the unknown polarization for $B^0\to\hc\Kstarz$ estimated as in~\cite{ref:Babar_chic2} ($6\%$). 

As a check, using the signal efficiency computed from MC events, the signal yield observed in data, and the number of \BB\ pairs in the data sample, we derive $\BR(B^+\to\etac\Kp)\times \BR(\etac\to K\bar{K}\pi)=(8.0\pm 0.4 (\mathrm{stat}))\times 10^{-5}$. This is in agreement with the world average value of $(6.4\pm 1.4)\times 10^{-5}$~\cite{ref:pdg2006}.

We calculate the ratios of the branching fractions with respect to $\BR(B^+\to\etac\Kp)$ using the ratios of signal yields and efficiencies with respect to $B^+\to\etac\Kp$, $R_{\Upsilon}=\Gamma(\FourS\to\B^+B^-)/\Gamma(\FourS\to B^0\bar{B}^0) = 1.026\pm 0.032$~\cite{ref:pdg2006}, and $\BR(\Kstarz\to \Kp\pi^-)=2/3$, and summing the uncertainties in quadrature. We define $R_{\etac K^*}=\BR(B^0\to\etac\Kstarz)/\BR(B^+\to\etac\Kp)$, $R_{\hc K}=\BR(B^+\to\hc\Kp)\times \BR(\hc\to\etac\gamma)/\BR(B^+\to\etac\Kp)$, $R_{\hc K^*}=\BR(B^0\to\hc \Kstarz)\times \BR(\hc\to\etac\gamma)/\BR(B^+\to\etac\Kp)$, $R_{\etactwos K}=\BR(B^+\to\etactwos K^{+})\times\BR(\etactwos\to\kkpi)/(\BR(B^+\to\etac\Kp)\times\BR(\etac\to\kkpi))$ and $R_{\etactwos K^*}=\BR(B^0\to\etactwos\Kstarz)/\BR(B^+\to\etactwos\Kp)$. Table~\ref{tab:systErr} summarizes the systematic uncertainties on the measurements. 

We obtain $R_{\etac K^*} = 0.62 \pm 0.06 (\mathrm{stat}) \pm 0.05 (\mathrm{syst})$, $R_{\etactwos K} = 0.096^{+0.020}_{-0.019} (\mathrm{stat})  \pm 0.025 (\mathrm{syst})$ and the $90\%$ C.L. upper limits $R_{\hc K} < 0.052$, $R_{\hc K^*} < 0.236$, and $R_{\etactwos K^*} < 1.0$. These are determined by assuming that each measurement follows a Gaussian distribution around the central value, with standard deviation given by the statistical and systematic uncertainties added in quadrature.
\begin{table}[!hbt]
  \caption{Summary of the relative contributions to the systematic errors on $R_{\etac K^*}$, $R_{\hc K}$, $R_{\hc K^*}$, $R_{\etactwos K}$ and $R_{\etactwos K^*}$.\\[0.5mm]}
  \begin{ruledtabular}
    \begin{tabular}{lccccc} 
      & \multicolumn{3}{c}{$\sigma(R)/R$ $(\%)$}  \\
      \rule[-2mm]{0mm}{4ex}&   $R_{\etac K^*}$ & $R_{\hc K}$ & $R_{\hc K^*}$ & $R_{\etactwos K}$ & $R_{\etactwos K^*}$\\
\hline
Signal yield                                  & 6.6 & 26 & 19 & 26 & 34\\
Signal efficiency                             & 1.4 &  2.2 &  6.7 & 1.3 & 2.2 \\
$R_{\Upsilon}$                                & 3.1 & $-$    & 3.1  & $-$ & 3.1\\
\hline
\rule[-1mm]{0mm}{4ex}Total                    & 7.2 & 26 & 20 & 26 & 34\\
    \end{tabular}
  \end{ruledtabular}
  \label{tab:systErr}
\end{table}

Using $\BR(B^+\to\etac\Kp) = (9.1\pm 1.3)\times 10^{-4}$, we derive 
\begin{equation}
\BR(B^0\to\etac\Kstarz) = (5.7\pm 0.6 (\mathrm{stat}) \pm 0.4 (\mathrm{syst}) \pm 0.8 (\mathrm{br}))\times 10^{-4},\nonumber
\end{equation}
where the last error is from the uncertainty on $\BR(B^+\to\etac\Kp)$, and the $90\%$ C.L. upper limits
\begin{eqnarray}
\BR(B^+\to\hc \Kp)\times \BR(\hc\to\etac\gamma) &<& 4.8\times 10^{-5},\nonumber \\
\BR(B^0\to\hc \Kstarz)\times \BR(\hc\to\etac\gamma) &<& 2.2\times 10^{-4}.\nonumber 
\end{eqnarray}

Using the world average value $\BR(B^+\to\etactwos K^{+})=(3.4\pm1.8)\times10^{-4}$\cite{ref:pdg2006}, we derive 
\begin{equation}
\BR(B^0\to\etactwos \Kstarz(890)) < 3.9\times10^{-4}, \nonumber
\end{equation}
at the $90\%$ C.L. Finally, using $\BR(B^+\to\etac\Kp)\times\BR(\etac\to\kkpi)=(6.88\pm0.77^{+0.55}_{-0.66})\times10^{-5}$~\cite{ref:Belle_etac}, we derive 
\begin{equation}
\BR(\etactwos\to\kkpi)=(1.9\pm0.4 (\mathrm{stat}) \pm0.5 (\mathrm{syst}) \pm1.0 (\mathrm{br}) )\%, \nonumber
\end{equation}
where the last error accounts for the uncertainties on the branching fractions used in the calculation. 

In summary, we obtain a measurement of $\BR(B^0\to\etac\Kstarz)$ in agreement with, and greatly improving upon, the previous world average value~\cite{ref:pdg2006}. We obtain an upper limit for $\BR(B^+\to\hc \Kp)\times \BR(\hc\to\etac\gamma)$ in agreement with the previous Belle result~\cite{ref:belle_exclhc}, and set the first upper limit on $\BR(B^0\to\hc \Kstarz)\times \BR(\hc\to\etac\gamma)$: these confirm suppression of \hc\ production in $B$ decays. We report the first upper limit on $\BR(B^0\to\etactwos \Kstarz)$ and the first measurement of $\BR(\etactwos\to\kkpi)$. The latter branching fraction is smaller than the corresponding branching fraction for \etac, and can be used to derive $\Gamma(\etactwos\to\gaga)$.  We measure $m(\etac)=2985.8\pm1.5\pm3.1~\mevcc$ and $\Gamma(\etac)=36.3^{+3.7}_{-3.6}\pm4.4~\mev$. These are in agreement with previous \babar\ measurements from $\gamma\gamma$ collisions~\cite{ref:Babar_etac} and slightly higher than the world average values~\cite{ref:pdg2006}.

\section{ACKNOWLEDGMENTS}
\label{sec:Acknowledgments}

\input pubboard/acknowledgements

\end{document}

%% file: definitions.tex
\def\mX {\ensuremath{m_X}}

\def\kskpi {\ensuremath{\KS \Kpm \pimp}}
\def\kkpi {\ensuremath{K \Kb \pi}}
\def\hc {\ensuremath{h_c}}
\def\etactwos {\ensuremath{\etac{(2S)}}}

%% file: pubboard/authors_feb2008.tex
%
\author{B.~Aubert}
\author{M.~Bona}
\author{Y.~Karyotakis}
\author{J.~P.~Lees}
\author{V.~Poireau}
\author{E.~Prencipe}
\author{X.~Prudent}
\author{V.~Tisserand}
\affiliation{Laboratoire de Physique des Particules, IN2P3/CNRS et Universit\'e de Savoie, F-74941 Annecy-Le-Vieux, France }
\author{J.~Garra~Tico}
\author{E.~Grauges}
\affiliation{Universitat de Barcelona, Facultat de Fisica, Departament ECM, E-08028 Barcelona, Spain }
% ...
\collaboration{The \babar\ Collaboration}
\noaffiliation

%% file: pubboard/acknowledgements.tex
We are grateful for the 
extraordinary contributions of our \pep2\ colleagues in
achieving the excellent luminosity and machine conditions
that have made this work possible.
The success of this project also relies critically on the 
expertise and dedication of the computing organizations that 
support \babar.
The collaborating institutions wish to thank 
SLAC for its support and the kind hospitality extended to them. 
This work is supported by the
US Department of Energy
and National Science Foundation, the
Natural Sciences and Engineering Research Council (Canada),
the Commissariat \`a l'Energie Atomique and
Institut National de Physique Nucl\'eaire et de Physique des Particules
(France), the
Bundesministerium f\"ur Bildung und Forschung and
Deutsche Forschungsgemeinschaft
(Germany), the
Istituto Nazionale di Fisica Nucleare (Italy),
the Foundation for Fundamental Research on Matter (The Netherlands),
the Research Council of Norway, the
Ministry of Education and Science of the Russian Federation, 
Ministerio de Educaci\'on y Ciencia (Spain), and the
Science and Technology Facilities Council (United Kingdom).
Individuals have received support from 
the Marie-Curie IEF program (European Union) and
the A. P. Sloan Foundation.

%% file: bad1808_la.bbl
\begin{thebibliography}{99}
\bibitem{ref:cc_note} Charge conjugated modes are implied throughout.
\bibitem{ref:hc_simul} In MC events, the total width of the \hc\ meson is $1~\mev$.
\end{thebibliography}
